# Where Is Information Management Research?

Tom Wilson and Elena Macevičiūtė

**Abstract.** We report on a preliminary investigation into the current scope of research in information management, adopting a conceptual approach derived from previous work by Hjørland in information science and by Palvia in information systems. We created a data-set of 107 articles resulting from a search in Web of Science, using the search strategy of the term *information management* in the titles of articles, and then restricting the analysis to those journals we identified as having an information science orientation, rather than an information systems orientation. The analysis reveals the *International Journal of Information Management* as the most significant journal in the field, but also draws attention to the rise of interest in the field through contributions to two Brazilian journals and one Spanish journal. The thematic analysis revealed that the dominant research themes from the information science perspective were *empirical user studies, studies of the structural and institutional approach,* and *information system usage and adoption.* Further work will be undertaken to explore the relevance of the approach in the analysis of other document sets from areas such as health care, construction and engineering.

## 1 Introduction

There have been previous attempts to identify the nature of *information management*; first, a paper that dealt with the creation of a curriculum for the subject [18], and then a further review [8], which was updated for publication in a collection of papers in 2004 [9], as well as a very computer-oriented analysis by Targowski [15]. Clearly, over the past fifteen years, many technological developments have taken place that are changing the nature of *information management*: cloud computing, big data, machine learning, social media, and further developments in mobile computing involving 4G and 5G and the worldwide spread of the smartphone. Over the same period, some of the earlier issues in information management will have assumed lesser significance and others will have been resolved. Consequently, our research objective here is to determine how the research field of information management has changed over the past fifteen years, in response to the changes that have taken place in the business and research environments. This paper is restricted to an analysis of the information-science-related papers on *information management* from journals that we identified as having an information science perspective on the subject. We intend to continue the analysis in the future by analyzing journals from other fields, in which information management articles can be found, e.g. health research journals.

There also have been calls to explore the information management research area more rigorously from the perspective of conceptual approaches [10; 17]. Articles exploring some conceptual approaches and features of information related fields and trying to find a common ground for them in general have been already published [3; 4; 11; 19]. These articles themselves have been written from different perspectives, e.g. looking for the essence of interdisciplinarity in library and information science [11], or building a model unifying essential conceptual elements of all information-related study areas [19], or identifying major study areas for uncovering the essence of information science [2]. Investigating research approaches in such a complex study area as information management seemed appropriate in the light of these previous developments. As we wanted to link our present investigation to the previous ones at least to some extent, we have formulated the following research questions:

1) Which journals publish articles on information management?
2) What conceptual approaches to exploring information management problems have been applied during the last 15 years in information science journals?
3) What topics are covered in the articles dedicated to information management research in information science journals?

We realize that answering all three research questions requires application of a rigorous theoretical and methodological means and a variety of research methods: content analysis, bibliometrics, and discourse analysis. In fact, we planned to conduct a rigorous domain analysis in information management research area [2], but covering all this in one conference paper may be quite impossible. Therefore, we are presenting here the results of the first explorative phase of a future wider investigation. We cover a limited number of articles in a clearly restricted group of journals and apply only content analysis to check the grounds. The concrete methods are presented in the methods part.

## 2    Theoretical approaches and main concepts

The starting point in the approach to this investigation is based on the idea proposed by Madsen [10] that there are "three underlying concepts: Information Management$_1$: at the institutional level; Information Management$_2$: content-oriented information management, rooted in information science; and Information Management$_3$: technology-oriented information management, rooted in information systems" (p. 536). Only the latter two are underpinned by conceptual approaches from two different disciplines, while the first consists of applications in institutions of various kinds and does not bring in any conceptual approach to the phenomena under study.

The idea seemed sound to us as even a glance at two online encyclopedias - Wikipedia and Citizendium - seem to correspond with the proposition. While Wikipedia presents a view on information management based mainly on information systems point of view, the Citizendium article reflects the information science approach. There are overlaps between the two with regard to the elements of organizational behaviour and business strategies, i.e., research areas that can be attributed to the institutional level perspectives and provide another dimension of interdisciplinarity of the information management research area [11].

Having to explore the output in information management from disciplinary perspectives we run into the problem of recognizing gold before we know what the gold is [5]. So, we have accepted the broad understanding of information science domain developed by Hjørland [2] and already applied to the analysis of information management area by Tirador Ramos [17]. In the analysis of the retrieved information management articles we apply eight out of 11 approaches of information science identified by [2]: 1) organizing information sources, 2) classifications and tesauri, 3) indexing and retrieval, 4) information user studies; 5) bibliometric studies; 6) organizing traditions and paradigms of information management; 7) document and genre studies; and 8) studies of structures and institution.

Of course, Hjørland's approaches pertain to the analysis of a particular research area and are meant to explore its structures. We are using them in a different way and treating them as types of studies that we may find in information management articles conducted from the information science perspective, e.g., creating classifications for organizational documents, organizing information sources for promotion, or conducting a user study for information audit in an organization. We have set aside epistemological and critical studies, terminological studies and professional cognition approaches as they will require more in-depth approach than could be required from an explorative study, but if needed we intend to use them too. We also do not expect to find all named approaches in the articles on information management, but rather to identify those that receive most attention in this research field.

We have derived the understanding of content structure in information systems research from two articles that have applied a method similar to ours to identify these structures in the journal *Information & Management*. The authors have examined the publications in this journal in 2007 [12] and 2017 [13]. In these articles, they among other things have established research topic trends in information systems journals [13, p. 225]. We began with the intention of applying the topic list in Table 12 of that paper, but, in performing the analysis discovered that mainly the following were applicable to the papers in this set: information systems design and development, information systems evaluation, social networks, security and privacy, and 'knowledge management' (here signifying mainly artificial intelligence, expert systems, and neural networks). We used other topic trends from the table when needed.

Hjørland and Palvia adopt different epistemological bases for their analyses of the different fields, but they allow us to identify the main conceptual approaches used in the retrieved sample of articles.

The institutional perspective is not well explained by Madsen [10]. So, we have defined it as the primary concern with any type of institution or organization (e.g. school, company, academic library) and investigation of information related problems in this organization. In fact, focus on "information in organization" was a primary criteria of article selection in both earlier studies. We also expect to find some mixed information science or information systems approach as well as some other additional perspectives in the articles with this primary concern.

To capture more elements pertaining to information management studies than just a particular approach we looked for some aids in creating a suitable coding scheme. We wanted an open scheme that would allow coding of various aspects of articles, but will also ensure structuring of the material and bringing in codes that we have not envisioned at the start of coding. Looking through a number of possibilities we have remembered the fundamental categories of the faceted classification by Ranganathan (1960) and realized that they meet our requirements. Thus we have used the following categories based on recent interpretation of them by Ferreira et al. [1]:

>**Personality** that in our case represented three approaches defined by Madsen and more detailed approaches identified in two conceptual approaches from information science and information systems disciplines.
>
>**Matter** consisting of objects, materials, and substances or data under exploration of the authors of the articles.
>
>**Energy** encompasses activities or processes of the researched subjects and methods and techniques applied by the investigators.
>
>**Space** mainly pertained to the place, in which a particular study was conducted, but also included some virtual spaces if they were clearly defined and essential for the given study.
>
>**Time**, on the other hand, was mainly limited to the publishing date of the article, but when the authors have specified the period of study, it was also included in the coding making clear distinction by the two types of time categories.

## 3  Method

The approach in the previous papers, cited above, was to select what appeared to be the key journals in the field of information management, on the basis of the number of relevant papers published. Thus, for 2002 and 2004 the journals *Information Economics and Policy, Information & Management, Information and Organization, International Journal of Information Management, Journal of Strategic Information Systems,* and *MIS Quarterly* were selected. Over the past fifteen years, however, and with only a cursory examination, some of these journals seem no longer to have a strong focus on information management and, in addition, new journals have arrived on the scene, with *information management* in the title. For example, *Aslib Proceedings* has changed its name to the *Aslib Journal of Information Management* and we now also have *Journal of Enterprise Information Management, Journal of Global Information Management,* and the *Journal of Information and Knowledge Management.*

For these reasons we decided to proceed by first selecting, through Web of Science, those journal articles published over the period that had *information management* in the title, removing *Lecture Notes in Computer Science* and *Lecture Notes in Artificial Intelligence,* which are book series, not journals. *Web of Science* produces a list of only the first 100 titles, ranked by the number of papers: forty-four of these journals had only two papers with information management in the title over the fifteen year period, and a further thirty journals had published only three papers in the period. If we were to choose only those journals that had published an average of one paper a year, i.e., fifteen papers in the period, we would have only two journals to consider: *International Journal of Information Management* (with 37 papers) and, rather curiously, *Anesthesia and Analgesia,* (with 19 papers). We examined the nineteen papers in this last journal and found that they were well spread over the period, rather than appearing in thematic issues in the journal, and almost all were concerned with the design, development, utilization and evaluation of anesthesia information management systems.

Clearly, to examine only these two journals as representative of the information management research area would be rather limiting. Consequently, we changed the search strategy to include papers with *information management* in the Topic field. This resulted in a list of thirty-six journals that had published at least one paper a year. The list is shown in the Appendix, divided into five categories: information science related – 13 journals; health and medicine related (including health informatics) – 13 journals; construction, engineering and technology related – 7 journals; general science (*PLOS One),* and business management *(MIS Quarterly Executive).*

The distribution of these 4,710 papers over the time period is shown in Figure 1. This is a rather odd distribution, with steady growth from 2007 until 2013, a sudden drop in 2014 and then quite massive growth (from 267 to 451 papers) in 2015, which was continued in the following years to 2018. If production continues at the present rate in 2019, the figure for 2018 (454 papers) will be equaled or exceeded.

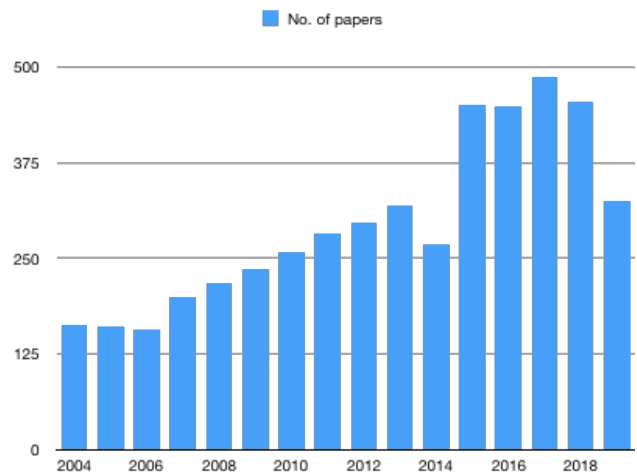

**Fig. 1.** Distribution of journal papers over time

There are several possible reasons for the sudden surge of publications from 2015 onwards: new journals may have been admitted to Web of Science in that year, but we find that only fourteen new titles were adopted [16]. When the papers for that year are examined, seventy-seven of the 100 journals listed carry only one or two papers tagged as *information management*, which perhaps suggests some sudden surge of interest across a wide range of disciplines. Once again, however, the journals with most papers are in the fields of information science and medicine and health-care. Whatever the reason for the surge, the graph shows that output has continued at approximately the same level since 2015.

There are several possible reasons for the sudden surge of publications from 2015 onwards: new journals may have been admitted to Web of Science in that year, but we find that only fourteen new titles were adopted [16]. When the papers for that year are examined, seventy-seven of the 100 journals listed carry only one or two papers tagged as *information management*, which perhaps suggests some sudden surge of interest across a wide range of disciplines. Once again, however, the journals with most papers are in the fields of information science and medicine and healthcare. Whatever the reason for the surge, the graph shows that output has continued at approximately the same level since 2015.

Scanning through some articles containing *information management* in the Topic field, we have discovered another possible reason for the surge and a potential problem for our investigation. We could not, in many cases, determine why the topic *information management* had been assigned to an article. For example, this title, retrieved in this manner, gives no indication of being centrally concerned with information management: "*Developing elementary students' digital literacy through augmented reality creation: insights from a longitudinal analysis of questionnaires, interviews, and projects*". Consequently, cleaning the data would have required significant time and effort. Thus, we have returned to the strategy of looking for *information management* in the title, in the belief that it would properly reveal the focus of research.

To compare the content of information management research papers with that revealed by the earlier studies we needed to select a manageable number of journals, no more than ten, which would give us a reasonably representative coverage of the fields of interest. However, this proved problematical, particularly when we tried to restrict the number of papers to examine by selecting a single year's output. As noted above, interest appears to have been aroused in a wide range of disciplines and this affects the distribution of papers over journals in any one year. For example, when we selected the latest complete year, i.e., 2018, this resulted in the retrieval of eighty-eight articles. However, only two journals had three articles with *information management* in the title in that year: *International Journal of Information Management* and *Anesthesia and Analgesia*, and only a further six journals had two papers. Only two of these six appear in the list of top journals in the Appendix, the *International Journal of Medical Informatics* and *Perspectivas em Ciencia da Informacão.*

Given this situation, we decided that our focus should be on the information science and information systems perspective. Madsen [10] has argued for the existence of conceptually different perspectives on information management: an information science perspective, and an information systems perspective. There is certainly some basis in the research literature for supporting this view and, therefore, it makes sense to focus on the information science perspective to identify the relations between the two approaches. Consequently, we decided to make the list of information-science-related journals in the Appendix, the focus of our further examination of the research outputs. That list, in rank order of the number of papers, consists of:

*International Journal of Information Management*
*Electronic Library*
*Journal of Documentation*

*Profesional de la Información*
*Aslib Proceedings*
*Informação Sociedade Estudos*
*Information Research: an International Electronic Journal*
*Perspectivas em Ciência da Informação*
*Library Hi Tech*
*Online Information Review*
*Records Management Journal*
*Journal of Information Science*
*Journal of The American Society For Information Science and Technology*
*Program. Electronic Library and Information Systems*

Two of the journals (*Aslib Proceedings* and *Journal of The American Society For Information Science and Technology*) changed their titles during the period and, therefore, papers in the new titles were also included).

A search using these titles, again with *information management* in the title, produced a total of 107 papers, with none being found for *Online Information Review* and the *Records Management Journal* (the items attributed to these journals were book reviews, not articles). The details of these papers were exported into an EndNote database, for further analysis.

It seemed rather surprising that no journals from the field of information systems appeared in the list. A search was run on Web of Science for papers with *information management* in the title, and *information systems* in the name of the publication. Again, this resulted in no records being found: however, capitalizing each word resulted in the retrieval of three conference papers, not journal articles. Investigating further, when the titles of journals in the field of information systems were used, articles were found. Thus, *information management* and *Information Systems Journal*, retrieved one paper; *European Journal of Information Systems* retrieved two; while *Journal of Strategic Information Systems, Information Systems Research,* and *Journal of Management Information Systems* produced nothing. So, we have focused on the journals in information science domain.

## 4   Findings

We are reporting our findings in this part of the article in relation to the research questions formulated at the beginning. Thus we explore the distribution of the journals and the locales, in which the studies were conducted, the presence and mix of conceptual approaches and distribution of topics over the journals.

### 4.1   Distribution of the Articles over Journals

Table 1 shows the distribution of papers over the journals. Probably the most interesting points to note in this respect are the significance of the *International Journal of Information Management* for research in the field, and the inclusion to two Brazilian journals, one Spanish journal, and one from Ecuador in the list. In the previous studies, these journals were not included in Web of Science and, as a result, escaped the notice of the authors. Their appearance here as representing key journals in their respective languages (Portuguese and Spanish) is testament to the significance of information management in these countries. It also suggests that researchers in the field would be well-advised to include these journals in their literature searches, since it is obvious that significant work is going on in these countries.

**Table 1.** Distribution of papers over journals.

| Journal title | No. | % |
| --- | --- | --- |
| International Journal of Information Management | 37 | 34.6 |
| Aslib Proceedings/Aslib Journal of Information Management | 10 | 9.3 |
| Perspectivas em Ciência da Informação | 10 | 9.3 |
| Informação Sociedade Estudos | 9 | 8.4 |
| Journal of the American Society/Association for Information Science and Technology | 8 | 7.5 |
| Profesional de la Información | 8 | 7.5 |
| Information Research | 6 | 5.6 |
| Library Hi Tech | 5 | 4.7 |
| Journal of Information Science | 4 | 3.7 |
| Revista Publicando | 4 | 3.7 |
| Program | 3 | 2.8 |
| Electronic Library | 1 | 0.9 |
| Journal of Documentation | 1 | 0.9 |
| Records Management Journal | 1 | 0.9 |

### 4.2 Distribution of Papers by Research Locale

Table 2 presents the distribution of papers by the area in which the research was carried out, or to which it is related. Thus, research may have been carried out in, say, Canada, but used an Internet-based site to collect data. Hence, the research locale was the Internet, rather than Canada. In addition to the countries and regions listed a further 13 that had one paper each were identified: i.e., Africa, Finland, Germany, Ghana, Israel, Kuwait, Malaysia, Mexico, Portugal, Singapore, Slovenia, Sweden, and the United Arab Emirates.

The table shows that the most common locale was designated *Global*: that is, the research was not restricted to any specific place, but the topic of research and the findings could apply to any location. These papers tended to be more theoretical and/or more information systems design-related. An analysis of the 26 papers from a conceptual perspective shows that 15 adopt an information science approach, six an information systems approach, and five an institutional perspective.

Table 2. Distribution of papers by research locale.

| Locale | No. | % | Cu % |
|---|---|---|---|
| Global | 26 | 24.3 | |
| Brasil | 15 | 14.0 | 38.3 |
| USA | 12 | 11.2 | 49.5 |
| Spain | 10 | 9.3 | 58.9 |
| UK | 10 | 9.3 | 68.2 |
| Australia | 4 | 3.7 | 72.0 |
| Canada | 3 | 2.8 | 74.8 |
| Cuba | 3 | 2.8 | 77.6 |
| Europe | 2 | 1.9 | 79.4 |
| Greece | 2 | 1.9 | 81.3 |
| Internet | 2 | 1.9 | 83.2 |
| Poland | 2 | 1.9 | 85.0 |
| South Africa | 2 | 1.9 | 85.0 |
| Norway | 2 | 1.9 | 86.9 |

Most of the papers distinctly connected to a particular territory or even an organization tend to be conducted in information science and on the institutional level. More than half of studies in both these categories are on the studies performed in particular locality.

### 4.3 Distribution by Conceptual Framework and Theme

Table 3 shows a thematic analysis of those papers that adopted an information science perspective. The themes in italics are additional to those suggested by Hjørland's domain analysis, and two (systems design and development, and systems evaluation) are drawn from Palvia's analysis of the information systems domain.

Table 3. Themes of papers with an information science framework.

| Theme | No. of papers |
|---|---|
| Empirical user studies. | 19 |
| Organising information sources. | 6 |
| Structures and institutions. | 6 |
| Bibliometric studies. | 4 |
| *Curriculum development.* | 3 |
| Indexing and retrieval. | 3 |
| Document and genre studies. | 1 |
| Epistemological and critical studies. | 1 |
| *Information policy* | 1 |
| *Relationship with e-learning* | 1 |
| Special classifications and thesauri. | 1 |
| TOTAL | 46 |

The most obvious point of interest here is that the dominant theme in information management issues within the information science framework is the *empirical user study*, which accounts for 41% of the forty-six papers. No other theme achieves more than thirteen per cent. Also of interest is the fact that we had to create three additional topics for education and curriculum, information policy, and e-learning.

Only eighteen papers fitted the information systems context and eight of these, perhaps not surprisingly, dealt with *information systems design and development*, or, in this context, the design and development of information management systems. As in the case of the information science approach, a category relating to *education* for information management had to be added and three papers dealt with this. Another three fall into the topic of system evaluation, while only one paper was found in the topics of *system usage and adoption*, *security and privacy*, and *social networks*. We also needed an extra topic of *data interpretation* for one of the papers in this set.

Table 4 shows the analysis of papers in the *institutional* focus: themes in the regular font are from Hjørland's domain analysis; those in italics are from Palvia's analysis of the information systems field, and the one item in bold italic was created to describe one paper.

**Table 4.** Themes of papers within the institutional focus

| Theme | No. | % | Cu % |
|---|---|---|---|
| Structures and institutions | 15 | 35.0 | |
| *Information usage and adoption* | 12 | 28.0 | 63 |
| Empirical user study | 7 | 16.0 | 79.0 |
| Organising information sources | 3 | 7.0 | 86.0 |
| *Electronic commerce* | 1 | 2.3 | 88.3 |
| Indexing and retrieval | 1 | 2.3 | 90.6 |
| *Information systems design and development* | 1 | 2.3 | 93.2 |
| ***Information systems education*** | 1 | 2.3 | 95.5 |
| *Security and privacy* | 1 | 2.3 | 97.8 |
| Traditions and paradigms | 1 | 2.3 | 100.1 |
| TOTAL | 43 | | |

**Note**: *Italics* show the topics from information systems perspective, ***italics and bold*** show a topic created additionally. The rounding of figures accounts for the result of more than 100 per cent.

The articles assigned to the category institutional were reporting studies, the focus of which was primarily a single organization, a group of organizations, inter-organizational relations or networks, and the research problem focused on information in a particular organizational setting. In addition to the themes we have also assigned to them the conceptual approach label showing whether the researchers had applied information science or information systems perspective. Most of the studies have applied a mixed information science and information systems perspective (15), an information science perspective was found in 10, and an information systems perspective in 13 articles.

It is also worth drawing attention to the presence of institutional context that was identified in the articles assigned primary conceptual approach tags. It was easily identified in 10 out of 13 information systems articles when examining e-commerce or management information systems problems or running case studies. It was also present in 35 (66%) of information science articles with highly dominant higher education and research institutions in student surveys or in relation to scholarly communication. As Table 5 shows, the biggest number of organisations, investigated by the authors, belong to the public sector (universities, research institutes, municipalities are among frequent). Private organisations are twice less common and small and medium size enterprises are heading this list. The inter-sectoral alliances are much less investigated and we have found only one article related to the activities of NGO in Brazil.

Table 5. Papers in relation to researched sectors.

| Sector | No. of papers | % |
| --- | --- | --- |
| Public | 48 | 44.9 |
| Private | 23 | 21.5 |
| Mixed | 4 | 3.7 |
| Non-governmental | 1 | 0.9 |
| Sector unrelated | 31 | 29.0 |

In relation to the identified approaches and topics it is necessary to mention a high number of articles on personal information management (20) that was found primarily under the information science approach sometimes mixed with information system requirement features.

### 4.4 Distribution of Themes over Journals

The journal with five or more papers with *information management* in the title were analysed by theme, with the results presented in Table 6. The result suggests that, from an information science perspective, the main themes within information management are empirical user studies, structures and institutions, and information systems usage and adoption. The latter theme is heavily influenced by the *International Journal of Information Management,* which published ten of the eleven articles. The three topics account for 56% of the 91 articles covered by the table, and the remaining 44% is accounted for by 15 topics.

Table 6. Frequency of specific themes over journals.

| Theme | IJIM | ASLIB | ISE | IR | JASIST | LHiT | PCI | PDIi | Total |
|---|---|---|---|---|---|---|---|---|---|
| Empirical user study | 6 | 6 | | 4 | 5 | | 2 | 1 | 24 |
| Structures and institutions | 5 | | 3 | | | | 5 | 3 | 16 |
| *Information usage and adoption* | 10 | | | | | 1 | | | 11 |
| *Information systems design and development* | 4 | 1 | | | 2 | | 1 | | 8 |
| Organising information sources | 2 | | 1 | | | 3 | | 2 | 8 |
| **Education and curriculum** | 2 | 1 | | 1 | | 1 | 1 | | 6 |
| *Systems evaluation* | 1 | | 1 | | | | | 2 | 4 |
| Indexing and retrieval | 2 | | | | 1 | | | | 3 |
| Bibliometric studies | | 1 | 2 | | | | | | 3 |
| *Security and privacy* | 2 | | | | | | | | 2 |
| *Social networks* | 1 | | | | | | | | 1 |
| *Electronic commerce* | 1 | | | | | | | | 1 |
| *Knowledge management* | 1 | | | | | | | | 1 |
| Traditions and paradigms | | | 1 | | | | | | 1 |
| Document and genre studies | | | 1 | | | | | | 1 |
| Classification and thesauri | | | | 1 | | | | | 1 |
| **Information policy** | | | | | | | 1 | | 1 |
| TOTAL | 37 | 10 | 9 | 6 | 8 | 5 | 10 | 8 | 93 |

## 5 Discussion and Conclusions

Overall, our analysis reveals the intricate interaction of the information science and information systems approaches to information management. There are clearly certain areas, such as the design and development of information management systems, where the information systems approach is dominant, and others, such as empirical user studies where the information science approach dominates.

Turning to our research questions, within the information science perspective of information management, we can identify quite clearly the journals that publish in the field (RQ1): quite obviously, the leading journal is the *International Journal of Information Management,* having published 34% of the papers with the term in the title of papers. It is followed by the *Aslib Journal of Information Management* (formerly *Aslib Proceedings*), and then by two Brazilian journals (*Informação Sociedade Estudos* and *Perspectivas em*

*Ciência da Informação*) and the Spanish journal *Profesional de la Información.* The inclusion of these journals in Web of Science since the papers by Wilson [1] (1989) and Maceviciute and Wilson [1] (2002, 2004), is a major change, drawing attention to the work now being carried out in this field in the Luso-Hispanic region.

We can also draw on the results of our preparatory process and identify three other groups of journals exhibiting clear interest in information science topics: 1) health and medicine, 2) bioinformatics and 3) construction, engineering and technology. It is worth exploring these publications in greater detail, especially in the light of meagre information management studies in the journals of information systems and business management.

Regarding the conceptual approaches to information management (RQ2), we had some difficulty in assigning papers to conceptual categories because of the close connection in some research between the information systems approach and the information science approach. However, on the basis of this rather limited data-set, we find that the information science approach and research on the institutional level dominate, with relatively few papers fitting the information systems conceptual framework as a dominating one. However, it may be subordinate to information science approach and even dominant in the studies of institutions and to our choice of journals.

A topic, or thematic analysis (RQ3) reveals that, within the information science journals, the empirical user study is the most common research topic, followed by papers on the structural and/or institutional framework for information management, and papers on information system usage and adoption.

Clearly, further work, especially the analysis of the health information and construction and engineering perspectives will reveal other journals of significance and other topics of relevance within those fields.

This has been a preliminary investigation of a limited set of papers relating to *information management,* partly to develop and test the method, and partly to find a basis upon which to conduct further investigation. The findings are limited by the size, and distribution over journals, and the conceptual framework employed. A different approach, for example, would have drawn attention to the emergence of *personal information management* as a concept in the field, which was almost completely lacking earlier. Almost 20% of the papers dealt with this topic from one perspective or another. Most of these papers were examples of empirical user studies.

Information management, from a research perspective, is a complex problem area: from an information science perspective its principal interest is in empirical research; from the information systems perspective its principal interest is as a subject of the design, development and application of information systems. From the perspectives of the health and medicine field and the construction and engineering fields, a cursory examination of papers suggests that the principal interest is in developing, using and evaluating systems that will improve practice in these areas.

We hope, in further work, to explore these differences and relationships more thoroughly.

## References

1. Ferreira, A.C., Maculan, B.C.M.d.S., Naves, M.M.L.: Ranganathan and the faceted classification theory. TransInformacao, Campinas, 29(3), 279-295 (2017)
2. Hjørland, B.: Domain analysis in information science: Eleven approaches - traditional as well as innovative. J Doc. 58(4), 422-462 (2002).
3. Hjørland, B.: Library and information science (LIS): Part 1. Knowl. Organ. 45(3), 232-254 (2018).
4. Hjørland, B.: Library and information science (LIS). Part 2. Knowl. Organ. 45(4), 319-338 (2018).
5. Hjørland, B.: Theories are knowledge organizing systems (KOS). Knowl. Organ. 42(2), 113-128 (2015).
6. Information management. Citizendium (2013, last edit) http://en.citizendium.org/wiki/Information_management, last accessed 2019/09/10
7. Information management. Wikipedia (2019, last edit) https://en.wikipedia.org/wiki/Information_management, last accessed 2019/09/10
8. Maceviciute, E., Wilson, T.D.: The development of the information management research area. Inform. Res. 7(3) (2002) http://InformationR.net/ir/7- 3/paper133.html last accessed 2019/09/10
9. Maceviciute, E., Wilson, T.D.: The development of the information management research area. In E. Maceviciute, T.D. Wilson, (Eds.). Introducing information management: an Information Research reader, p. 18-30. London: Facet Publishing (2004).
10. Madsen, D.: Disciplinary perspectives on information management. Procedia – Soc. Behav. Sci. 73, 534-537 (2013).
11. Madsen, D.: Liberating interdisciplinarity from myth: An exploration of the discursive construction of identities in information studies. J Assoc. Inf. Sci. Tech. 67(11), 2697-2709 (2015).
12. Palvia, P., Pinjani, P., Sibley, E.H.: A profile of information systems research published in *Information & Management.* Inform. Manage. 44(1), 1-11 (2007).
13. Palvia, P., Chau, P.Y.K., Kakhki, M.D., Choshal, T., Uppala, V., Wang, W.: A decade plus long introspection of research published in *Information & Management.* Inform. Manage. 54(2), 218-227 (2017).
14. Ranganathan, S.R.: Colon classisfication: basic classification. Bombay: Madras Library Association Publication (1960).
15. Targowski, A.: A definition of information management discipline. In: Khosrowpour, M. (ed.). Information-Resources-Management-Association International Conference, Vancouver, Canada, 1997, pp. 301-307. Hershey, PA: Idea Group Publishing.
16. Testa, J.: A view from Web of Science: journals, articles, impact. Information Services & Use 36(1-2), 99-104 (2016), https://content.iospress.com/articles/information-services-and-use/isu801, last accessed 2019/08/28
17. Tirador Ramos, J.: El dominio e su implicación para la gestión de la información. Investigatión Bibliotecologica 24(50), 49-60 (2010)/
18. Zhang, P.; Benjamin, R.I.: Understanding information related fields: A conceptual framework. J Assoc. Inf. Sci. Tech. 58(13), 1934-1947 (2007).
19. Wilson, T.D.: Towards an information management curriculum. J. Inform. Sci. 15(4-5), 203-209 (1989).

**Appendix:** Journals publishing articles with "information management" as a key-word in the title, key-words and abstract sorted under the Web of Science subject areas and listing number of retrieved articles in the brackets

**Information science**
International Journal of Information Management (107)
Electronic Library (43)
Journal of Documentation (42)
Profesional de la Informacion (40)
Aslib Proceedings (35)
Informacao Sociedade Estudos (33)
Information Research: an International Electronic Journal (27)
Perspectivas em Ciencia da Informacao (27)
Library Hi Tech (23)
Online Information Review (23)
Records Management Journal (22)
Journal of Information Science (21)
Journal of The American Society For Information Science and Technology (18)
Program Electronic Library and Information Systems (16)
**Health and medicine**
Anesthesia and Analgesia (62)
Telemedicine and E Health (57)
Health Information Management Journal (50)
International Journal of Medical Informatics (47)
Journal of The American Medical Informatics Association (32)
BMC Bioinformatics (30)
Journal of Medical Systems (27)
Methods of Information In Medicine (20)
Journal of Clinical Monitoring and Computing (18)
Journal of Medical Internet Research (18)
Anesthesiology (16)
Applied Clinical Informatics (16)
CIN Computers Informatics Nursing (16)
**Construction, engineering and technology**
**Automation in Construction (31)**
Journal of Computing in Civil Engineering (28)
Journal of Construction Engineering and Management (28)
Industrial Management Data Systems (21)
Advanced Engineering Informatics (15)
Agro Food Industry Hi Tech (15)
IEEE Access (15)
**General science**
PLOS One (16)
**Business management**
MIS Quarterly Executive (15)